\documentclass[a4paper,12pt]{article}
\usepackage{amsmath,amsfonts,amssymb,cite}

\begin{document}

\begin{center}
{\Large\bf No-wall holographic model for QCD}
\end{center}

\begin{center}
{\large S. S. Afonin}
\end{center}

\begin{center}
{\it V. A. Fock Department of Theoretical Physics, Saint-Petersburg
State University, 1 ul. Ulyanovskaya, 198504, Russia}
\end{center}

\begin{abstract}
We propose a new type of bottom-up holographic model describing the
Regge like spectrum of mesons. This type of spectrum emerges due to
condensation of a scalar field in the bulk anti-de Sitter space. The
gauge invariance of the action describing the vector mesons requires
that the bulk scalar field must correspond to an operator of
canonical dimension two. This is not necessary for the description
of the scalar sector where the generation of the mass spectrum can
be directly related to the formation of gluon condensate in QCD. It
is also demonstrated that such an explicit introduction of local
dim2 operator into the model leads to disappearance of the
corresponding contribution from the dim2 condensate in the power
expansion of the gauge-invariant vector correlator as required by
QCD.
\end{abstract}

\section{Introduction}

The bottom-up holographic models inspired by the ideas of AdS/CFT
correspondence~\cite{witten,gub} turned out to be an interesting
phenomenological approach to the physics of strong interactions
encompassing in a uniform framework the low-energy effective field
theories and the QCD sum rules in the large-$N_c$ limit. The
simplicity of theoretical setup represents a nice feature of
holographic approach to QCD. At the same time, the setup of many
AdS/QCD models looks {\it ad hoc} and motivated by the phenomenology
rather than by some consistent logic. In spite of extensive
literature on this subject appeared in the last few years (an
incomplete list is given by
Refs.~\cite{son1,pom,son2,andreev,braga3,hirn2,maru,cs,
shock2,batell,fazio3,forkel2b,vega2,gh,
afoninP,afoninP2,br2,co2,zuo,redi,afoninI}) some important questions
about the foundations of holographic models remain unanswered. One
of these questions is the origin and physical meaning of the dilaton
background in the action of the so-called soft-wall models which
were introduced in Ref.~\cite{son2,andreev} to describe
holographically the Regge-like spectrum. The corresponding
background $e^{\pm \Lambda^2z^2}$ (we will refer to such soft-wall
models as SW$^{\pm}$) violates explicitly the Lorentz invariance
along the fifth coordinate $z$ and it looks therefore puzzling how
it could arise from a consistent solution of five-dimensional
Einstein equations (see, however, the toy model of
Ref.~\cite{batell}). On the other hand, this background provides the
Regge-like spectrum for the radially excited states,
$m_n^2\sim\Lambda^2n$, introducing simultaneously the mass scale
$\Lambda$. In the present work, we will try to shed light on the
problem by transforming the SW model into a holographic model with a
dynamical mechanism for mass generation based on condensation of a
scalar field. Such a "no-wall model" turns out to differ from the
original SW model representing rather a mean of the SW$^+$ and
SW$^-$ models. This property leads to interesting consequences, for
instance, the dimension-two condensate in the power expansion of
vector correlator is absent automatically in the no-wall model.

The paper is organized as follows. The no-wall holographic model is
introduced in Section~2 . In Section~3, we calculate the two-point
vector correlator and clarify the emerging difference between the SW
and no-wall models. Our conclusions are summarized in Section~4.

\section{The no-wall model from the soft-wall model for vector mesons}

Consider the simplest action of the SW$^{\pm}$ model for vector
mesons without the chiral symmetry breaking term,
\begin{equation}
\label{1}
S=\int d^4\!x\,dz\sqrt{g}\,e^{\pm \Lambda^2z^2}\left\{
-\frac{1}{4g_5^2}F_{MN}F^{MN}\right\},
\end{equation}
where $F_{MN}=\partial_M V_N-\partial_N V_M$, $M=0,1,2,3,4$. One
writes usually the action~\eqref{1} in terms of the left and right
fields, $F_L^2+F_R^2$, this will be not essential for us, so the
parity of $V_M$ is not fixed at this stage. For simplicity, we do
not introduce also the isospin index, i.e. the action~\eqref{1} is
supposed to describe the isosinglet vector states, say, the
$\omega$-mesons. Our analysis, however, will be valid for all light
vector mesons ($\rho$, $a_1$ {\it etc.}) since we are interested in
the mass spectrum, hence, only the quadratic terms in~\eqref{1} are
relevant. The coupling $g_5$ plays the role of normalization
constant for the 5D vector field $V_M$ and can be fixed after
calculation of the two-point vector correlator (see
relation~\eqref{24} below). The UV boundary value $V_M(x,0)$ is
interpreted as the source for a QCD operator interpolating the
vector mesons, e.g., for $\bar{q}\gamma_{\mu}q$ if we consider the
$\omega$-mesons. The metric is the usual AdS one,
\begin{equation}
\label{2}
ds^2=\frac{R^2}{z^2}(dx_{\mu}dx^{\mu}-dz^2),\qquad z>0,
\end{equation}
here $R$ is the radius of the AdS space and $z$ is the holographic
coordinate. According to the prescriptions of the AdS/CFT
correspondence~\cite{witten,gub}, the masses of fields in the dual
theory defined on the AdS$_5$ space are
\begin{equation}
\label{3}
m_5^2R^2=(\Delta-J)(\Delta+J-4),
\end{equation}
where $\Delta$ is the canonical dimension of the corresponding 4D
field theory operator and $J$ is the spin. In our case, the field
$V_M$ is massless. This field is considered as the gauge one, so we
may use the gauge freedom to fix the axial gauge, $V_z=0$, which is
implied in what follows.

The Lorentz invariance along the $z$-direction is explicitly
violated in the action~\eqref{1} due to the dilaton background. Let
us remove this background with the help of the transformation
\begin{equation}
\label{4}
V_M=e^{\mp \Lambda^2z^2/2}\tilde{V}_M.
\end{equation}
The action becomes equivalent to
\begin{equation}
\label{5} S=\int d^4\!x\,dz\sqrt{g}\left\{
-\frac{1}{4g_5^2}\tilde{F}_{MN}\tilde{F}^{MN}+\frac{\Lambda^4z^4}{2R^2g_5^2}\tilde{V}_M\tilde{V}^M\right\},
\end{equation}
where instead of the $z$-dependent background we have now a
$z$-dependent mass term. The action~\eqref{5} can be regarded as the
definition of new holographic model formulated in terms of the field
$\tilde{V}_M$ (we will omit tildes in what follows). The
$z$-dependent mass contribution breaks the invariance with respect
to the local gauge transformations of the field $\tilde{V}_M$ that
corresponds to violation of a global flavor symmetry in the
equivalent 4D field theory together with the breaking of the scale
invariance ($z$ is the inverse energy scale). There is, however, a
way to formulate the model in a gauge-invariant manner: The
appearance of the $z$-dependent term may be interpreted as an effect
of condensation of some bulk scalar field which is coupled to the
vector field via the covariant derivative,
\begin{equation}
\label{6}
S=\int
d^4\!x\,dz\sqrt{g}\left\{|D_M\varphi|^2-m_{\varphi}^2\varphi^2
-\frac{1}{4g_5^2}F_{MN}F^{MN}\right\},
\end{equation}
where $D_M=\partial_M-iV_M$. Actually, the action~\eqref{6} is in
one-to-one correspondence with the simplest holographic setup
describing the chiral symmetry breaking~\cite{son1,pom}. The
difference consists in the interpretation of the bulk scalar field
$\varphi$. The equation of motion for $\varphi$ in the absence of
the vector field $V_M$ is
\begin{equation}
\label{7}
-\partial_z\left(\frac{\partial_z
\varphi}{z^3}\right)+\frac{m_{\varphi}^2R^2\varphi}{z^5}=0.
\end{equation}
In order to have the same equation of motion for $V_M$ that would
follow from the action~\eqref{5} the Eq.~\eqref{7} must lead to the
solution $\varphi_0\sim z^2$. This dictates $m_{\varphi}^2R^2=-4$.
The relation~\eqref{3} tells us that the field $\varphi$ corresponds
then to the operator of canonical dimension $\Delta=2$. Such a
gauge-invariant local operator does not exist in QCD. Nevertheless,
there are many suggestions in the literature~\cite{dim2} that the
corresponding gluon operator is required in the phenomenology for an
effective parametrization of various non-perturbative effects.

It should be mentioned that if we describe the non-gauge fields the
introduction of dim2 operator is not necessary. For instance, one
may write the following "no-wall model" for the scalar mesons
\begin{equation}
\label{7b}
S_{\text{scal}}=\int
d^4\!x\,dz\sqrt{g}\left\{\left(\partial_M S\right)^2-m_s^2S^2
-\lambda\tilde{\varphi}S^2+\left(\partial_M\tilde{\varphi}\right)^2-m_{\tilde{\varphi}}^2\tilde{\varphi}^2\right\}.
\end{equation}
To have the Regge-like spectrum from the equation of motion for $S$
the field $\tilde{\varphi}$ must have the bulk expectation value
$\tilde{\varphi}_0\sim z^4$. According to the equation of
motion~\eqref{7} this is achieved if $m_{\varphi}^2R^2=0$ and for
the canonical dimension of $\tilde{\varphi}$ we obtain
from~\eqref{3} the value $\Delta=4$. One can construct only one
operator with $\Delta=4$ in the chiral limit:
$G_{\mu\nu}^aG^{\mu\nu,a}$. In QCD, it acquires a vacuum expectation
value  --- the gluon condensate $\langle G^2\rangle$ --- that causes
the scale anomaly in the trace of energy-momentum tensor. Since the
fifth coordinate $z$ has the physical meaning of inverse energy
scale, the condensation of the scalar field $\tilde{\varphi}$ is
dual to the emergence of non-trivial gluon condensate in QCD ---
both effects make different energy scales non-equivalent. The fact
that the anomaly is a quantum effect while in the holographic model
the breaking of the scale invariance looks spontaneous should not be
surprising as long as the dual theory is supposed to describe {\it
semiclassically} the quantum effects in the strong coupling regime.
Note also that the field $\tilde{\varphi}$ is in one-to-one
correspondence with the dilaton field considered in the original
paper~\cite{witten} for demonstration of the idea of AdS/CFT
correspondence.

According to an AdS/CFT prescription~\cite{kleb}, the solution of
classical equation of motion for a scalar field $\Phi$ corresponding
to an operator $O$ has the following form near the 4D boundary
$z\rightarrow0$,
\begin{equation}
\label{10}
\Phi(x,z)\rightarrow
z^{4-\Delta}\left[\Phi_0(x)+\mathcal{O}(z^2)\right]+z^{\Delta}\left[\frac{\langle
O(x)\rangle}{2\Delta-4}+\mathcal{O}(z^2)\right],
\end{equation}
where $\Phi_0(x)$ acts as a source for $O(x)$ and $\langle
O(x)\rangle$ denotes the corresponding condensate. Thus one has
$\Lambda^4\sim\langle G^2\rangle$ in the case of the
model~\eqref{7b}. The exact solution for $\tilde{\varphi}_0$,
\begin{equation}
\label{8}
\tilde{\varphi}_0(z)=C_1+C_2z^4,
\end{equation}
has the form~\eqref{10}. The constant $C_2$ dictates the slope of
the scalar trajectory, $C_2\lambda R^2=\Lambda^4$, while the
constant term $\lambda C_1$ renormalizes the 5D mass $m_s$ in the
action~\eqref{7b}. It is curious to note that the dim2 condensate is
not a well defined object in the relation~\eqref{10}.

\section{Mass spectrum and matching to OPE}

The mass spectrum of vector particles can be calculated following
the standard procedure~\cite{son1} by solving the equation of motion
for the transverse components of the field $V_\mu$ in the
action~\eqref{6} or~\eqref{5} (the gauge $V_z=0$ is implied),
\begin{equation}
\label{12a}
\left[-\frac{q^2V_\mu(q,z)}{z}-\partial_z\left(\frac{\partial_z
V_\mu(q,z)}{z}\right)+\Lambda^4zV_\mu(q,z) \right]_{\bot}=0,
\end{equation}
where $V_\mu(q,z)$ is the 4D Fourier transform of $V_\mu(x,z)$. By
assumption, the physical particles correspond to the normalizable
solutions of Eq.~\eqref{12a}. According to the Sturm-Liouville
theorem, such solutions $v_n(z)$ exist only for discrete values of
4D momentum $q^2$ which are identified with the on-shell masses
$m_n^2$,
\begin{equation}
\label{12b} -\partial_z\left(z^{-1 }\partial_z v_n\right)+\Lambda^4z
v_n=z^{-1 } m_n^2 v_n.
\end{equation}
The corresponding solutions are
\begin{equation}
\label{12c}
m_n^2=4\Lambda^2(n+1),
\end{equation}
\begin{equation}
\label{12d}
v_n(z)=\sqrt{\frac{2n!}{(n+1)!}}e^{-z^2\Lambda^2/2}z^2\Lambda^2
L_n^1(z^2\Lambda^2).
\end{equation}
Here $L_n^1$ denotes an associated Laguerre polynomial. The
spectrum~\eqref{12c} coincides with the spectrum of the SW$^\pm$
models but the exponential factor in the eigen-function~\eqref{12d}
differs.

The alternative way for finding the spectrum is to calculate the
vector two-point correlator and find its poles. We will show this
calculation in detail in order to demonstrate the arising difference
between the no-wall model and the SW one.

The two-point correlation function of vector currents $J_{\mu}$ is
defined as
\begin{equation}
\label{13} \int d^4x e^{iqx}\langle
J_{\mu}(x)J_{\nu}(0)\rangle=(q_{\mu}q_{\nu}-q^2g_{\mu\nu})\Pi_V(Q^2),\qquad
Q^2=-q^2.
\end{equation}
According to the AdS/CFT prescription~\cite{witten,gub}, the vector
two-point correlator is given by the following relation~\cite{son1},
\begin{equation}
\label{14}
\Pi_V(Q^2)=-\frac{R}{g_5^2Q^2}\left[V(Q,z)\frac{\partial_z
V(Q,z)}{z}\right]_{z\rightarrow 0},
\end{equation}
where $V(Q,z)$ stems from the 4D Fourier transform
$V^{M}(Q,z)=V(Q,z)V^{M}_0(Q)$ for the 4D transverse part of $V^{M}$
and can be found from the equation of motion,
\begin{equation}
\label{15} -\partial_z\left(\frac{\partial_z
V}{z}\right)+\Lambda^4zV=-\frac{Q^2}{z}V,
\end{equation}
with the boundary conditions,
\begin{equation}
\label{16}
V(Q,0)=1,\qquad V(Q,\infty)=0.
\end{equation}
The solution is
\begin{equation}
\label{17}
V(Q,z)=\Gamma\left(1+\frac{Q^2}{4\Lambda^2}\right)e^{-z^2\Lambda^2/2}U\left(\frac{Q^2}{4\Lambda^2},0;z^2\Lambda^2\right),
\end{equation}
where $U$ is the Tricomi confluent hypergeometric function. This
solution differs from the corresponding expression in the SW model
by the exponential factor: In the SW$^-$ model, this factor is
absent while in the SW$^+$ one the factor is $e^{-z^2\Lambda^2}$.
Thus, the vector "bulk-to-boundary propagator" $V(Q,z)$ of the
no-wall holographic model represents the logarithmic mean of the
corresponding propagators of SW$^-$ and SW$^+$ models. The same
property takes place for the wave functions of 4D discrete
modes~\eqref{12d}. Near the boundary, the solution~\eqref{17} has
the following expansion,
\begin{multline}
\label{18} V(Q,z)_{z\rightarrow
0}=\left(1-\frac{z^2\Lambda^2}{2}+\mathcal{O}(z^4\Lambda^4)\right)\times\\
\left\{1+\frac{Q^2}{4\Lambda^2}\left[\log(z^2\Lambda^2)
+\psi\left(1+\frac{Q^2}{4\Lambda^2}\right)+2\gamma-1\right]z^2\Lambda^2+\mathcal{O}(z^4\Lambda^4)\right\}=\\
1+\left\{\frac{Q^2}{4\Lambda^2}\left[\log(z^2\Lambda^2)
+\psi\left(1+\frac{Q^2}{4\Lambda^2}\right)+2\gamma-1\right]-\frac12\right\}z^2\Lambda^2+\mathcal{O}(z^4\Lambda^4),
\end{multline}
where the expression in the first line emerges from the exponential
factor in~\eqref{17}. The Digamma function can be represented as
\begin{equation}
\label{19}
\psi\left(1+\frac{Q^2}{4\Lambda^2}\right)=\sum_{k=1}^{\infty}\frac{1}{k}-
\sum_{k=1}^{\infty}\frac{1}{\frac{Q^2}{4\Lambda^2}+k}-\gamma.
\end{equation}
Substituting now the expansion~\eqref{18} into the
relation~\eqref{14} and exploiting the representation~\eqref{19},
after subtracting the contact terms and redefining $k=n+1$ we arrive
at our final answer for the vector correlator,
\begin{equation}
\label{20}
\Pi_V(Q^2)=\frac{R}{2g_5^2}\left[\sum_{n=0}^{\infty}\frac{4\Lambda^2}{Q^2+4\Lambda^2(n+1)}+\frac{2\Lambda^2}{Q^2}\right].
\end{equation}
The poles $Q^2_n=-m_n^2$ in the sum of expression~\eqref{20} yield
the mass spectrum~\eqref{12c}. As was said above, this spectrum
coincides with the spectrum of SW$^{\pm}$ model for the vector
mesons.

The massless pole that is present in the correlator turns out to be
indispensable for providing the correct high energy asymptotic
behavior. Making use of the expansion
\begin{equation}
\label{21}
\psi\left(1+\frac{Q^2}{4\Lambda^2}\right)_{Q^2\rightarrow\infty}=
\log\left(\frac{Q^2}{4\Lambda^2}\right)+\frac{2\Lambda^2}{Q^2}-\frac{4\Lambda^4}{3Q^4}+
\mathcal{O}\left(\frac{\Lambda^8}{Q^8}\right),
\end{equation}
we obtain the following asymptotic expansion of the correlator under
consideration,
\begin{equation}
\label{22}
\Pi_V(Q^2)_{Q^2\rightarrow\infty}=\frac{R}{2g_5^2}\left[\log\left(\frac{4\Lambda^2}{Q^2}\right)+\frac{4\Lambda^4}{3Q^4}+
\mathcal{O}\left(\frac{\Lambda^8}{Q^8}\right)\right].
\end{equation}
The expansion~\eqref{22} can be matched to the Operator Product
Expansion (OPE) for the vector two-point correlator~\cite{svz},
\begin{equation}
\label{23}
\Pi_V(Q^2)_{Q^2\rightarrow\infty}=\frac{N_c}{24\pi^2}\log\left(\frac{\mu_{\text{ren}}^2}{Q^2}\right)+
\frac{\alpha_s}{24\pi}\frac{\langle G^2\rangle}{Q^4}+
\mathcal{O}\left(\frac{\mu_{\text{ren}}^6}{Q^6}\right).
\end{equation}
The matching of coefficients  in front of the logarithmic terms
yields the standard normalization factor for the 5D vector field,
\begin{equation}
\label{24}
\frac{R}{g_5^2}=\frac{N_c}{12\pi^2}.
\end{equation}
An important feature of the no-wall model is the absence of
$\mathcal{O}(Q^{-2})$ contribution in the expansion~\eqref{22} as
required by the OPE~\eqref{23} --- in QCD in the chiral limit, there
is no local gauge invariant condensate of dimension two. The
cancelation of $\mathcal{O}(Q^{-2})$ contribution occurs due to the
massless pole in the correlator~\eqref{20}. Looking at the
expansion~\eqref{18} it is easy to notice that the pole term stems
from the exponential factor. As follows from the remark after the
relation~\eqref{17} the pole term is absent in the SW$^-$ model and
is twice the pole term of~\eqref{20} in the SW$^+$ model. Thus, both
the SW$^-$ and the SW$^+$ models possess a feature that we do not
expect in QCD --- they predict a non-zero dimension-two condensate
in the OPE,
\begin{equation}
\label{24a}
\Pi_V(Q^2)^{\text{SW}^\pm}_{Q^2\rightarrow\infty}=
\frac{R}{2g_5^2}\left[\log\left(\frac{4\Lambda^2}{Q^2}\right)\pm\frac{2\Lambda^2}{Q^2}+
\mathcal{O}\left(Q^{-4}\right)\right].
\end{equation}
The dimension-two condensate of the no-wall model represents the
arithmetic mean of the dimension-two condensates of the SW$^+$ and
the SW$^-$ models and as seen from~\eqref{24a} this very property
cancels the $\mathcal{O}(Q^{-2})$ contribution.

\section{Conclusions}

We have proposed a new type of bottom-up holographic model
describing the Regge-like spectrum of the radially excited states.
In contrast to the soft wall models, the considered setup does not
contain the dilaton background. The introduced dynamical mechanism
for mass generation is not novel --- it is widely exploited for
description of the chiral symmetry breaking in the holographic
approach. The spontaneous chiral symmetry breaking contributes to
the masses of low-lying states but it is definitely not dominant in
the generation of masses of highly excited resonances. Within the
presented scheme, a similar mechanism provides the main contribution
to the masses of all light mesons from a dual description of
gluodynamics. The distinguishing features of the no-wall model with
respect to the soft wall one are: The physical interpretation is
more transparent (e.g., the appearance of discrete meson spectrum
and its Regge form are directly related to the formation of gluon
condensates in QCD), the absence of dimension-two condensate in the
power expansion of the two-point correlators, the model is closer to
the original AdS/CFT theoretical setup (the pure AdS space in the
bulk), and the exploited Higgs-like mechanism for mass generation
looks more natural for the field-theoretical description.

The most important lesson of the present work is that the effective
presence of dim2 operator cannot be removed from the holographic
models describing the Regge like spectrum: If we replace the soft
wall model by the no-wall one and remove thereby the dim2 condensate
from the power expansion of the vector correlators then this
contribution shows up directly in the action in the form of a scalar
field corresponding to the dim2 operator.

The scalar fields which generates the mass spectrum in the no-wall
model correspond to the gluon operators. Since the gluon sector is
involved one should analyze the back reaction of the metric in
response to the condensation of these fields. The given work is left
for future.

The properties of the no-wall holographic model suggest that this
model represents an alternative possibility for constructing
five-dimensional dual descriptions for QCD.

\section*{Acknowledgments}

I am grateful to my colleagues from the Department of Theoretical
Physics of Saint-Petersburg State University for fruitful
discussions. The work is supported by RFBR, grant 09-02-00073-a, by
SPbSU grant 11.0.64.2010, by the Alexander von Humboldt Foundation
(Return Fellowship) and by the Dynasty Foundation.


\begin{thebibliography}{99}
\bibitem{witten} E. Witten, Adv. Theor. Math. Phys. {\bf 2}, 253 (1998).
\bibitem{gub} S.~S.~Gubser, I.~R.~Klebanov and A.~M.~Polyakov, Phys. Lett. B {\bf 428}, 105 (1998).
\bibitem{son1} J.~Erlich, E.~Katz, D.~T.~Son and M.~A.~Stephanov,
Phys. Rev. Lett. {\bf 95}, 261602 (2005).
\bibitem{pom} L.~Da Rold and A.~Pomarol,
Nucl. Phys. B {\bf 721}, 79 (2005).
\bibitem{son2} A.~Karch, E.~Katz, D.~T.~Son and M.~A.~Stephanov,
Phys. Rev. D {\bf 74}, 015005 (2006).
\bibitem{andreev} O. Andreev, Phys. Rev. D~{\bf 73}, 107901 (2006).
\bibitem{braga3} H.~Boschi-Filho, N.~R.~F.~Braga and H.~L.~Carrion,
Phys. Rev. D {\bf 73}, 047901 (2006).
\bibitem{hirn2} J. Hirn, N. Rius and V. Sanz,
Phys. Rev. D {\bf 73}, 085005 (2006).
\bibitem{maru} K.~Ghoroku, N.~Maru, M.~Tachibana and M.~Yahiro,
Phys. Lett. B {\bf 633}, 602 (2006).
\bibitem{cs} C. Cs\'{a}ki and M. Reece, JHEP {\bf 0705}, 062 (2007).
\bibitem{shock2} J.~P.~Shock, F.~Wu, Y.-L.~Wu and Z.-F.~Xie, JHEP {\bf 0703}, 064 (2007).
\bibitem{batell} B. Batell and T. Gherghetta, Phys. Rev. D {\bf 78}, 026002 (2008).
\bibitem{fazio3} P.~Colangelo, F.~De Fazio, F.~Giannuzzi, F.~Jugeau and S.~Nicotri,
Phys. Rev. D {\bf 78}, 055009 (2008).
\bibitem{forkel2b} W.~de Paula, T.~Frederico, H.~Forkel and M.~Beyer,  Phys. Rev. D {\bf 79}, 075019 (2009).
\bibitem{vega2} A.~Vega and I.~Schmidt, Phys. Rev. D {\bf 79}, 055003 (2009).
\bibitem{gh} T. Gherghetta, J. I. Kapusta and T. M. Kelley,
Phys. Rev. D {\bf 79}, 076003 (2009).
\bibitem{afoninP} S.~S.~Afonin, Phys. Lett. B {\bf 675}, 54 (2009).
\bibitem{afoninP2} S.~S.~Afonin, Phys. Lett. B {\bf 678}, 477 (2009).
\bibitem{br2} G.~F.~de Teramond and S.~J.~Brodsky,
Phys. Rev. Lett. {\bf 102}, 081601 (2009).
\bibitem{co2} A.~Cherman, T.~D.~Cohen and E.~S.~Werbos,
Phys. Rev. C {\bf 79}, 045203 (2009).
\bibitem{zuo} F. Zuo, Phys. Rev. D {\bf 82}, 086011 (2010).
\bibitem{redi} D. Becciolini, M. Redi and A. Wulzer, JHEP {\bf 1001} (2010) 074.
\bibitem{afoninI} S.~S.~Afonin, Int. J. Mod. Phys. A {\bf 25}, 5683 (2010).
\bibitem{dim2} F. V. Gubarev and V. I. Zakharov, Phys. Lett. B {\bf 501}, 28 (2001);
K. G. Chetirkin, S. Narison, and V. I.~Zakharov, Nucl. Phys. B {\bf
550}, 353 (1999); K.-I. Kondo, Phys. Lett. B {\bf 514}, 335 (2001);
H. Verschelde, K. Knecht, K. Van Acoleyen, and M. Vanderkelen, Phys.
Lett. B {\bf 516}, 307 (2001); P. Boucaud et al., Phys. Rev. D {\bf
63}, 114003 (2001); E. Ruiz Arriola, P. O. Bowman, and W.
Broniowski, Phys. Rev. D {\bf 70}, 097505 (2004); E. Megias, E. Ruiz
Arriola, and L. L. Salcedo, JHEP {\bf 01}, 073 (2006); V. I.
Zakharov, hep-ph/0509114; S. Narison, hep-ph/0508259; E.~Ruiz
Arriola and W.~Broniowski, Phys. Rev. D {\bf 73}, 097502 (2006);
{\bf 81}, 054009 (2010).
\bibitem{kleb} I.~R.~Klebanov and E.~Witten, Nucl. Phys. B {\bf 556}, 89 (1999).
\bibitem{svz} M. A.~Shifman, A. I.~Vainstein and V. I.~Zakharov,
Nucl. Phys. B {\bf 147}, 385, 448 (1979).
\end{thebibliography}
\end{document}